\documentclass[final, oneeqnum,onetabnum,onealgnum]{siamart220329}

\ifpdf
\hypersetup{
pdftitle={Automating Variational Differentiation},
pdfauthor={K. Li and A. Damle}
}
\fi

\usepackage[T1]{fontenc}
\usepackage[ttdefault=true]{AnonymousPro}
\usepackage{listings}
\usepackage[utf8]{inputenc}
\usepackage{xcolor}
\usepackage[nosort]{cite}
\crefname{lstlisting}{listing}{listings}

\definecolor{lightgray}{rgb}{.96,.96,.96}
\definecolor{darkgray}{rgb}{.4,.4,.4}
\definecolor{purple}{rgb}{0.65, 0.12, 0.82}

\lstdefinelanguage{Julia}{
  keywords={typeof, true, false, begin, end, return, sum, @pct, @space, @domain},
  keywordstyle=\color{black}\bfseries,
  ndkeywords={export},
  ndkeywordstyle=\color{darkgray}\bfseries,
  identifierstyle=\color{black},
  comment=[l]{\#},
  commentstyle=\color{purple}\ttfamily,
  stringstyle=\color{red}\ttfamily,
  morestring=[b]"
}

\usepackage{amsmath}
\usepackage{amssymb}
\usepackage{amsfonts}

\title{Automating Variational Differentiation 
}
\author{Kangbo Li\thanks{Department of Computer Science, Cornell University (\email{kl935@cornell.edu})}
\and Anil Damle\thanks{Department of Computer Science, Cornell University (\email{damle@cornell.edu})}}
\headers{Automating Variational Differentiation}{K. Li and A. Damle}

\begin{document}
\maketitle
\begin{abstract}
 Many problems in Physics and Chemistry are formulated as the minimization of a
functional. Therefore, methods for solving these problems typically require
differentiating maps whose input and/or output are functions---commonly
referred to as variational differentiation. Such maps are not addressed at the
mathematical level by the chain rule, which underlies modern symbolic and
algorithmic differentiation (AD) systems. Although there are algorithmic
solutions such as tracing and reverse accumulation, they do not provide human
readability and introduce strict programming constraints that bottleneck performance,
especially in high-performance computing (HPC) environments.  In this manuscript,
we propose a new computer theoretic model of differentiation by combining the
pullback of the $\mathbf{B}$ and $\mathbf{C}$ combinators from the combinatory
logic. Unlike frameworks based on the chain rule, this model differentiates a
minimal complete basis for the space of computable functions. Consequently, 
the model is capable of analytic backpropagation and variational
differentiation while supporting complex numbers. To demonstrate the generality
of this approach we build a system named \texttt{CombDiff}, which can
differentiate nontrivial variational problems such as Hartree-Fock (HF) theory and multilayer perceptrons.
\end{abstract}



\section{Introduction}
\label{sec:Introduction}


Variational differentiation (VD)~\cite{CalculusOfVarGelfa2000} is the
differentiation of maps from functions to functions as well as maps from tensors
to tensors, which we will refer to as function maps and tensor maps. VD is
widely used in the physical sciences and applied mathematics. A classic
example is the principle of minimal action~\cite{ClassicalMechaHerber2002}, where
one differentiates the action of a particle $S(r) = \int \mathcal{L}(r(t),
\dot{r}(t)) \mathrm{d}t$ with respect to the trajectory $r(t)$---yielding
Newton's law for a classical particle. In discretized form, VD becomes
the differentiation of tensor maps. An example is to take the gradient of
a quadratic loss function $x^{*} Ax$, producing $2 A x$ if $A$ is
Hermitian.


A more challenging example that is representative of the type of differentiation
we aim to automate is the Hartree-Fock (HF)
energy~\cite{ComputationalPThijss2012, ASimplificatioSlater1951}, which is central to a
foundational method in electronic structure theory~\cite{MathematicalInLinL2019,
ElectronicStruMartin}. Within HF theory the $N_e$
electrons associated with a molecule can be approximately encoded as a matrix $C\in
\mathbb{C}^{N_b \times N_e}$ with orthogonal columns (i.e., $C^{*} C = I$),
where $N_b$ is the number of basis functions used to describe each
electron. The energy of the system then contains a term of the form 
\begin{equation}
\Sigma_{i j}^{N_e} \Sigma_{p q r s}^{N_b} C_{p i}^{*} C_{r j}^{*} J_{p q r
s} C_{s j} C_{q i},\label{eq:hartree-fock-j}
\end{equation}
where $J$ is a fourth order tensor that encodes the interaction between
electrons and has the (time reversal) symmetries $J_{pqrs} = J_{rspq} = J_{qpsr}^*$. To build
computational methods that minimize the energy, we need to differentiate
\cref{eq:hartree-fock-j} with respect to $C$. Moreover, when doing so it is
important to take into account the symmetries as they generally reduce the
evaluation cost of the gradient by a significant factor (essentially free in the case
of \cref{eq:hartree-fock-j}). In this paper, we aim to provide a  model for
automating VD and backpropgation analytically. 
In addition to producing simplified and human readable results, our approach prevents the
differentiation from interfering with numerical computation, which significantly 
complicates the implementation of differentiation systems while compromising the performance.

To accomplish this, we present a new pair of differentiation rules  for the
so-called $\mathbf{B}$ and $\mathbf{C}$ combinators from  combinatory logic.
Our rules serve as an alternative to combining the chain rule with  algorithmic
mechanisms such as partial evaluation and reverse accumulation. Avoiding 
algorithmic mechanisms enables the differentiation system to
\textit{analytically} backpropagate and perform general VD (even without the
discretization).  Moreover,  our approach makes it possible to optimize for
computational efficiency \textit{after} the gradient is analytically known,
thus separating differentiability constraints from performance concerns.


The advantage of our
approach is rooted in a foundational completeness result from the combinatory
logic, which states the $\mathbf{B}$ and $\mathbf{C}$ combinators\footnote{or equivalently, the $\mathbf{S}$ and $\mathbf{K}$ combinator.} form a minimal
complete basis for expressing all computable functions. These combinators and
the completeness result were first proposed a century ago by 
Schr\"ofinkel~\cite{UberDieBausteSchonf1924} and Curry~\cite{curry1930}. The
primary application of the combinators has been to develop the theory of
computability~\cite{curry1958combinatory} as an alternative approach to the
more popular lambda calculus~\cite{ASetOfPostulChurch1932}, but they have also
found practical use in building parsers~\cite{fokker1995functional,
leijen2001parsec}, reasoning about data updates~\cite{CombinatorsForFoster2007},
and automatic parallelization~\cite{InteractionComLafont1997}. Ultimately, our
approach scaffolds a (computational) framework for differentiation that is capable of VD and
backpropagation, preserves human readability, and enables performance optimization. We use this framework to build the \texttt{CombDiff} system.

\subsection{Relation to existing symbolic and computational systems systems}
Differentating~\eqref{eq:hartree-fock-j} is difficult for existing symbolic differentiation systems such as
SymPy~\cite{meurer2017sympy} and Mathematica, which are typically based on the
chain rule and a collection of primitive functions. Nevertheless, symbolic systems can
differentiate through univariate functions as well as some ``matrix calculus.''
This analytical approach has attractive advantages:
the result is human readable; the gradient is fast to evaluate; and the system is
simple to implement. However, there are several drawbacks that
severely limit its applicability. The first is that univariate functions
paired with some matrix calculus can only express a limited subset of the
tensor maps and does not include \cref{eq:hartree-fock-j}. The second is
that symbolic systems typically do not properly address complex numbers in the context of
optimization. Lastly, such methods do not support backpropagation---a necessity for a broad class of problems including machine learning
training and ab-initio molecular dynamics.

These problems can be remedied by algorithmic differentiation (AD) systems, but not without hefty compromises.
Mathematically, AD systems are also based on the chain rule\footnote{It is noteworthy that $\mathbf{B}$ is essentially
the composition rule. From this perspective, the chain rule misses the entire
dimension of functions spanned by $\mathbf{C}$, which is a problem that is
partly compensated for by allowing mutations. Moreover,
the chain rule only differentiates a composition $f_1(f_2(x))$ with
respect to the input $x$ but not functions $f_1$ and $f_2$, so it requires partial evaluations.} and
a collection of optimized numerical kernels~\cite{AutomaticDiffeBaydin2015, TheSimpleEsseElliot2018,
EvaluatingDeriGriewa2008}, which alone cannot enable
backpropagation. The additional ingredient that enables backpropagation and distinguishes AD systems from
symbolic systems is partial evaluation, which often appears in the form of
a computation graph traced at runtime~\cite{DonTUnrollAdInnes2018}. The
price of the partial evaluation is steep, with the obvious loss being
readability and symbolic processing power. The more subtle downstream
impact is in high performance computing (HPC) settings, where one has to develop performance optimizations without seeing the gradient. 
Furthermore, the differentiability constraints~\cite{jax2018github} allow only 
pure user functions, which exclude primitives indispensable to HPC such as mutations and communication.
As a result, applying basic HPC methods such as pipelining and recomputation to a problem require
major dedicated efforts~\cite{GpipeEfficienHuang2018, OptimalGradienFeng2018},
and sophisticated or problem-specific techniques such as leveraging tensor
symmetries and tensor compressions are still open questions. On the
positive side, relinquishing readability enables AD systems to
differentiate through complex numbers by treating them as 2D real vectors---a solution that is less acceptable for symbolic systems.

There are two partial solutions in AD for addressing tensor maps. The first is
to trade immutability and optimized kernels for generality by differentiating
low-level mutating statements~\cite{DonTUnrollAdInnes2018}. Tampering with the
immutability constraint can significantly complicate program
analysis~\cite{jaxmut}, cause bugs~\cite{tensoropmut}, and incur severe
performance loss due to the inability to use optimized
kernels~\cite{michael_abbott_2023_10035615}. A second strategy is to express
the tensor maps as a long sequence of matrix and vector operations. This approach is
limited to a subset of tensor maps and the performance is sub-optimal compared to tensor
contraction engines~\cite{cutensor, TensorProcessiGeorga2021,
	TheItensorSofFishma2020, TensorContractHirata2003,
michael_abbott_2023_10035615}. The overhead arises from creating an intermediate copy of the
tensors for each operation in the sequence, so it is inherent to the method and cannot be easily eliminated. 
Lastly, it is possible to add higher order functions to the set of primitives
of an AD system, which covers a few key types of functionals, but the lack of completeness 
is a problem~\cite{AutomaticFunctLinM2023}.

\section{Notation}

We will employ an ``anonymous'' notation for tensors and functions, which is less
commonly used but is necessary to compactly explain our developments. For example, the Fourier transform can be
written in terms of an anonymous function as 
\begin{equation}
f \mapsto (\omega \mapsto \textstyle\int \exp(-i\omega r) f(r) \mathrm{d} r),
\end{equation}
which maps $f$ to another anonymous function that represents the transformed
function $\hat{f}(\omega)$. One can apply an anonymous function to an input,
which itself can be an anonymous function, and the function application can be
evaluated by substitution. For example, we can apply the Fourier transform to an inverse Fourier transform
\begin{align}
    &\left(f\mapsto (\omega \mapsto \textstyle\int \exp(-i \omega r) f(r) \mathrm{d} 
    r)\right)\left(r \mapsto \textstyle\int \exp(i \omega r) g(\omega) \mathrm{d} \omega\right)\\
    =&\omega \mapsto \textstyle\iint \exp(-i\omega r) \exp(i \omega' r) g(\omega') 
    \mathrm{d} \omega' \mathrm{d} r \\
    =&\omega \mapsto \textstyle\int \delta(\omega - \omega') g(\omega')\mathrm{d} \omega'
    = g(\omega).
\end{align}
For our theory to be applicable to tensors, we adopt the same notation and treat
tensors as maps from integer indices to the corresponding tensor elements.  For
example, a vector $v$ can be written as $i \mapsto v_i$, which represents a
vector whose $i^{\mathrm{th}}$ element is $v_i$. We will also use the function evaluation notation $v(i)$ interchangeably with $v_i$ for
accessing tensor elements.


\section{Main Result} We present a pair of differentiation rules as an alternative
to the current technology stack for (computational) differentiation, primarily for automating
VD. As a demonstration and verification of our framework, we provide an
experimental system called
\texttt{CombDiff}\footnote{\url{https://github.com/kangboli/CombDiff}} that can
perform backpropagation and VD purely symbolically, and
produce a human readable gradient. Moreover, it can manipulate the result into
the appropriate computational kernels and simplify the result based on tensor
symmetries. 
We illustrate the efficacy of our system by applying it to
a variety of objective functions including quadratic
loss, conjugate gradient~\cite{MatrixComputatGolub2013,
MethodsOfConjHesten1952}, the HF energy, and the Maximally Localized Wannier
Function (MLWF) spread metric~\cite{AccuratePolariStenge2006,
MaximallyLocalMarzar1997}. 
In our examples, the symmetry-based simplification produces a gradient that can be evaluated for
free as an intermediate of the objective function. This cost reduction appears to 
be associated with the time reversal symmetry of the models.
The documentation page of \texttt{CombDiff} contains
more sample problems, including analytic backpropagation for a multilayer
perceptron.

As an example, the tensor contraction in \cref{eq:hartree-fock-j} can be
expressed and differentiated with the code in \cref{lst:hf-result}, a detailed
explanation of which is deferred to \cref{sec:Results} and the output is given
in
\cref{eq:hf-gradient}
\begin{lstlisting}[caption=Hartree Fock,label=lst:hf-result]
f, _ = @pct _ ctx (J::T) -> pullback((C::CM) -> sum((i, j, p, q, r, s),
    C(p, i)' * C(q, i) * C(r, j)' * C(s, j) * J(p, q, r, s))); f
\end{lstlisting}

\section{Background}%
\label{sec:Background}

In this section, we motivate the concept of the pullbacks and the chain rule
starting from how some scientists approach VD by hand,
from which we introduce pullbacks. We then explain \textit{composition
with captures} and the chain rule of pullbacks and go through an
example of obtaining derivatives using the formalism.

A common recipe for performing VD is based on ``first
order perturbation'' heuristics: the input of the function is perturbed and we observe the first-order change in the output. For example, when taking the trace of a matrix product
$(A, B) \mapsto \mathrm{tr}(A B)$ one can perform VD using the matrix product
rules and the trace rule
\begin{align}
    \delta \mathrm{tr}(X) &= \mathrm{tr}(\delta X), \quad
    \delta AB = (\delta A) B + A \delta(B),\label{eq:var-diff-rules}\\
    \delta \mathrm{tr}(AB) &= \mathrm{tr}((\delta A) B) + \mathrm{tr}(A (\delta B)) = \langle \delta A, B \rangle + \langle A, \delta B \rangle, \label{eq:trace-gradient}
\end{align}
from which one concludes that $B = 0$ and $A = 0$ are the stationary points.
This approach  is not suited for symbolic processing because it is hard to
express the idea of ``evaluating the derivative at a point'' without ambiguities. The systematic way to encodes the same
information in the AD theory is the \textit{pushforward}. We will
introduce the pushfowards and pullbacks in our notations; see, e.g.,~\cite{AutomaticDiffeBaydin2015,
EvaluatingDeriGriewa2008} for a review.



The concept of a \textit{pushforward} can be motivated by the Jacobian chain
rule for computing the gradient of a composite function $g(x) = f_1(f_2(x))$,
where $x$ is a vector,  $f_2: \mathbb{R}^M \to \mathbb{R}^N$ and $f_1:
\mathbb{R}^N \to \mathbb{R}$. The gradient of $g$ is 
\begin{equation}\label{eq:jacobian-product}
    \nabla g(x) = \mathcal{J}g(x)^T = \mathcal{J}f_2(x)^T \cdot \mathcal{J}f_1(f_2(x))^T,
\end{equation}
which is sufficient for deriving the gradient. However, the Joabian is often
$O(N)$ sparse and highly structured, so the $O(N^2)$ time and memory for forming
and multiply the Jacobian is unnecessary. This structured sparsity can be
captured and leveraged using the pullbacks $\mathcal{P}$ and
pushforwards $\mathcal{F}$, which are the building blocks of AD systems and can
be defined as   
\begin{align}
    \mathcal{P}f(x, k) \triangleq \mathcal{J} f(x)^T \cdot k
    &= i \mapsto \Sigma_j k_j \partial f(x)_j / \partial x_i\label{eq:real-pullback}, \\
    \mathcal{F}f(x, k) \triangleq k \cdot \mathcal{J} f(x)^T 
    &= i \mapsto \Sigma_j k_j \partial f(x)_i / \partial x_j\label{eq:real-forward}, 
\end{align}
both of which encode the Jacobian through its action (as a linear map)
instead of its entries.  A second reason why the pullback/pushforward are preferable 
to the Jacobian in our context is that they do not rely on the partial
derivative notation of the type $\partial f(x)_i/\partial x_j$, which is incompatible
with our formulation.\footnote{The difficulty is that $x_j$ is supposed to be a
independent variable but $j$ is not independent.}

To see the connection between the pushforwards and VD rules in  \cref{eq:var-diff-rules},
one can examine the pushforward of the matrix product following the definition \cref{eq:real-forward}
\begin{align}
        &\mathcal{F}((A, B) \mapsto AB) (A ,B, K, L) \\
        =& (p, q) \mapsto \Sigma_{i, j} K_{i j} \frac{\partial \Sigma_n A_{pn} B_{nq}}{\partial A_{ij}} 
         + \Sigma_{i,j} L_{i j}  \frac{\partial \Sigma_n A_{pn} B_{nq}}{\partial B_{ij}}\\
        =& (p, q) \mapsto \Sigma_n K_{pn} B_{nq} + \Sigma_n A_{pn} L_{nq} = K B + A L.
\end{align}
Identifying $K = \delta A$ and $L = \delta B$, this equation
corresponds to the product rule in~\cref{eq:var-diff-rules}. For the rest of
this manuscript, we choose to formulate our theory in terms of pullbacks because
it is aligned with reverse-mode AD systems, which are more suited for scientific optimization problems.

Given a collection of pullbacks, the chain rule based differentiation is
essentially the construction of the pullback of a function composition from
the pullbacks of its constituents. The formula
can be derived from rewriting~\cref{eq:jacobian-product} in terms of pullbacks.
\begin{align}
    \mathcal{P} g(x, k) = \mathcal{J} f_2(x)^T \cdot (\mathcal{J} f_1(f_2(x))^T \cdot k)
    = \mathcal{P}f_2(x, \mathcal{P}f_1 (f_2(x), k)) \label{eq:pullback-chain}.
\end{align}
The gradient can then be computed as $\nabla g(x) = \mathcal{P}(g)(x, 1)$. A
basic chain rule based differentiation system can then be viewed as a two-stage
process. First, one breaks down a function into \emph{a composition with captures}. The
captures are to be understood as the constants in a function. For example,
consider the composition $\exp \circ (x \mapsto w \cdot x)$. The weight $w$ is
what one would consider a ``constant,'' and these constants are called captures
in the programming language literature; formally they have to be treated as
variables declared in an outer scope. Thus, the full context of our function
would be a composition with a capture
\begin{equation}
w \mapsto \exp \circ (x \mapsto w \cdot x). \label{eqn:currying}
\end{equation}
The second stage is to derive the pullback of the composition using the chain rule
\begin{align}
    & w \mapsto \mathcal{P}(\exp \circ (x \mapsto w \cdot x)) \\
    =& w \mapsto (x, k) \mapsto 
    \mathcal{P}(x \mapsto w \cdot x)(x, \mathcal{P}(\exp)(w \cdot x, k))\\
    =& w \mapsto (x, k) \mapsto ((x, k) \mapsto w k)(x, \mathcal{P}((x, k) \mapsto \exp(x) k)(w \cdot x, k))\\
    =& w \mapsto (x, k) \mapsto w \exp(w \cdot x) k,
\end{align}
where the pullback of the multiplication and exponential are assumed given. 
The gradient is then obtained as $w \exp(w \cdot x)$ by setting $k = 1$.



\section{Model}%
\label{sec:Model}

In the section, we introduce the $\mathbf{B}$ and $\mathbf{C}$ combinators, then
we contribute the differentiation rules for the combinators
and a procedure for applying them to VD. Heuristically speaking, the
$\mathbf{C}$
combinator converts a function that returns a function to a composition with
captures.  For tensor problems, it converts the differentiation
of tensor maps to the differentiation of parametrized univariate functions.
The differentiation rule for the $\mathbf{B}$ combinator yields two terms: one
term serves the role of the chain rule, and the other term does the
backpropagation. We will not discuss the derivation of the rules, but we will
show how they work by walking through examples and verifying the correctness
through implementation. Lastly, we show that complex numbers can
be treated by extending the definition of pullback, and the $\mathbf{B}$ and
$\mathbf{C}$ rules still apply.

\subsection{The Combinators}%
\label{sub:The C combinator}

In combinatory logic, the composition of two functions $f_1$ and $f_2$ can be
modeled with the $\mathbf{B}$ combinator as $\mathbf{B}(f_1)(f_2)
= (\mathbf{B}(f_1))(f_2)$, where the function evaluation is left associative and
the $\mathbf{B}$ combinator is defined as
\begin{equation}
    \mathbf{B} = f \mapsto (g \mapsto (x \mapsto f(g(x)))).
\end{equation}
From this perspective, the AD process can be viewed as decomposing a function
in terms of the $\mathbf{B}$ combinator and a collection of primitive functions.
The composition can be differentiated by assembling the pullbacks of the
primitive functions using the chain rule. The drawback of this approach is that
it cannot backpropagate or express more general functions such as functions that return functions.

To differentiate functions that return functions, we need 
the $\mathbf{C}$ combinator, which is defined as
\begin{align}
    \mathbf{C} &= f \mapsto (x \mapsto (y \mapsto f(y)(x))).
\end{align}
To see what $\mathbf{C}$ does, we rewrite
a function returning a function in terms of $\mathbf{C}$
\begin{equation}
w \mapsto (x \mapsto \exp(w \cdot x)) = \mathbf{C}(x \mapsto 
\mathbf{B}(\exp) (w \mapsto w \cdot x))),\label{eqn:c-example}
\end{equation}
where the function inside $\mathbf{C}$ is a composition with a capture $x$---a
case we know how to differentiate from \cref{eqn:currying}. In abstract,
the $\mathbf{C}$ combinator converts a function that returns a function into a
composition with a captures. More generally, the process of decomposing a function into
the combinators and a small number of primitives, which are listed in
\cref{tab:primitive-lambdas}, is known as abstraction elimination.
\begin{table}[ht]
\begin{center}
\begin{tabular}{|l|c|c|}
    \hline
    primitive & definition & pullback\\\hline
    $\mathbf{B}$-combinator & $f \mapsto g \mapsto x \mapsto f(g(x))$ & See Eq.~\cref{eq:b-rule}\\
    $\mathbf{C}$-combinator & $g \mapsto x \mapsto b \mapsto g(b)(x)$ & See Eq.~\cref{eq:c-rule} \\\hline
    conjugate & $x \mapsto x^*$ & $(x, k) \mapsto k^*$\\
    multiply & $x \mapsto v x$ & $(x, k) \mapsto v^* k$ \\
    add & $x \mapsto v + x$ & $(x, k) \mapsto k$\\
    contract & $\Sigma$ & $(x, k) \mapsto i \mapsto k$\\
    \hline
\end{tabular}
\label{tab:primitive-lambdas}
\end{center}
\caption{The primitive functions and their pullbacks.}
\end{table}

\subsection{Differentiating the Combinators}%
\label{sub:The B combinator with captures}

We have reduced a function that return functions to the $\mathbf{C}$ combinator
and a composition with captures. To differentiate the $\mathbf{C}$
combinator, we need a new rule call the $\mathbf{C}$ rule, which can be written in
terms of the pullbacks as 
\begin{align}
    \mathcal{P} \left(\mathbf{C}(g)\right) = (x, k) \mapsto \Sigma( b \mapsto \mathcal{P}\left( g(b) \right)(x,  k(b))), \label{eq:c-rule}
\end{align}
where $\Sigma$ denotes a polymorphic contraction. Polymorphic means that its
definition depends on the type of its input. In the case of our rule
\cref{eq:c-rule}, if $b$ is a real number or a function, then $\Sigma$ is an
integral over the real line or the corresponding function space. We will use the
notation $\Sigma_b g(b)$ as a shorthand for $\Sigma(b \to g(b))$ to be familiar
and concise, but conceptually $\Sigma$ maps a function to a number. As an
example of the $\mathbf{C}$ rule, applying it to the previous example in
\cref{eqn:c-example} gives
\begin{align}
        &\mathcal{P}(\mathbf{C}(x \mapsto \mathbf{B}(\exp)(w \mapsto w \cdot x))(x, k)\\
        =& \Sigma_b \mathcal{P}(x \mapsto \mathbf{B}(\exp)(x\mapsto x \cdot b)(x))(x, k(b))\\
        =& \Sigma_b b \exp(b \cdot x) k(b),
\end{align}
The right hand side involves an inner product between $k$ and $b \mapsto b\exp(b \cdot
x)$, which seems unclear how to compute. However, the contraction generally does not 
need to be computed. Instead, $k$ typically turns out to contain a polymorphic delta
function that meets the contraction due to the $\mathbf{B}$ rule that we will
introduce next. The two polymorphic objects then subtly combine into a simple
substitution, an example of which will be shown in \cref{eq:product-rule}.

The differentiation rules for the $\mathbf{B}$ combinator may appear to be just
\cref{eq:pullback-chain}.  However, this does not account for the crucial
case where $x$ is captured by $f$ or $g$, which 
is required for the backpropagation. For
example, $f \mapsto f(i)$ can be written as $f \mapsto \mathbf{B}(f)
(\mathbf{I}) (i)$, where $\mathbf{I}$ is the identity map and the first input of
$\mathbf{B}$ is dependent on the input variable $f$. This necessitates an
adaptation to the chain rule, which we refer to as the $\mathbf{B}$ rule
\begin{equation}\label{eq:b-rule}
\begin{aligned}
        \mathcal{P}&\left( x \mapsto \mathbf{B}(f)(g)(x) \right) = (x, k) \mapsto \\
        & \mathcal{P}\left( x \mapsto g(x)\right) (x, \mathcal{P}\left( f \right)(g(x), k)) + \mathcal{P}\left( x \mapsto f\right) (x, i \mapsto \delta(g(x), i, k)),
\end{aligned}
\end{equation}
where $\delta$ is a polymorphic delta function similar to the polymorphic
contraction. $\delta(i, j, k)$ is $k$ if $i = j$ but $0$ otherwise. If $i, j$
are integers, real numbers, or functions, $\delta$ represents a Kronecker delta,
a Dirac delta, or a delta function on the function space. We have chosen to
write delta functions as function calls instead of superscripts/subscripts to
avoid abusing the notation for the Dirac/Kronecker delta. The first term in the
rule reproduces the chain rule, and the second term is responsible for
backpropagating, which includes a function returning a function $x \mapsto f$, necessitating
the $\mathbf{C}$ combinator. The delta function will meet the contraction that
appears when applying the $\mathbf{C}$ rule from \cref{eq:c-rule} and will
result in a substitution.

\subsection{Examples}%

To illustrate the significance of 
the backpropagation part of the $\mathbf{B}$ rule, we consider differentiating the solution of a system of equations $g(x) = 0$. When
$g$ arises from a gradient $g = \nabla R$ this corresponds to differentiating 
a stationary point of $R$, which models a broad class of problems including
molecular dynamics and sensitivity analysis\cite{fiacco1990nonlinear}. Consider
the procedure $\rho$ that maps $g$ to one of its roots, where the condition $g
\mapsto g(\rho(g)) = g \mapsto 0$ is satisfied. Differentiating both sides yields
\begin{align}
    0 
    &= \mathcal{P}(g \mapsto \rho(g))(g, \mathcal{P}g (\rho(g), k))
    + \mathcal{P}(g \mapsto g)(g, i \mapsto \delta(i, \rho(g), k))\\
    &= \mathcal{P}(\rho)(g, H(k)) + i \mapsto \delta(i, \rho(g), k),
    \quad H(k) = \mathcal{P}g(\rho(g), k)\label{eq:meet}\\
    \mathcal{P}(\rho) &=(g, k) \mapsto i \mapsto \delta(i, \rho(g), -H^{-1}(k)),
\end{align}
where $H(k)$ is a linear map on $k$ that represents multiplying by the Hessian
of $R$ if $g = \nabla R$. This example essentially differentiates the function $\mathrm{argmin}$~\cite{Gould2016OnDP}, which is typically performed
after $g$ is concretely known and $g(x) = 0$ has been solved.


The backpropagating part of the $\mathbf{B}$ rule also enables us to  decompose
any explicit function in terms of only univariate functions. For example, $x
\mapsto f(x) \cdot g(x)$ can be decomposed using only multiplication by a constant
as $x \mapsto (v\mapsto f(x)\cdot
v)(x)$ because the $(v \mapsto f(x) \cdot v)$ is allowed to capture $x$. This
implies that multivariate calculus of static dimensions is already handled
through the $\mathbf{B}$ rule and that multivariate primitives such as adding or
multiplying two numbers are unnecessary. The differentiation of $x \mapsto
f(x)\cdot g(x)$ proceeds as
\begin{align}
    &\mathcal{P}(x \mapsto (v \mapsto f(x) \cdot v)(g(x))) (x, k)\label{eq:product-rule}\\
    =& \mathcal{P}(g)(x, \mathcal{P}(v \mapsto f(x) \cdot v)(g(x), k)) \\
     &+ \mathcal{P}(x \mapsto (v \mapsto f(x) \cdot v))(x, i \mapsto \delta(i, g(x), k))\\
    =& \mathcal{P}(g)(x, f(x)^* k) + \Sigma_v \mathcal{P}(x \mapsto (t \mapsto t\cdot v)(f(x))) (x,  \delta(v, g(x), k))\\
    =& \mathcal{P}(g)(x, f(x)^* k) + \Sigma_v \mathcal{P}(f)(x, \mathcal{P}(t \mapsto t \cdot v)(f(x), \delta(v, g(x), k)))\label{eq:contraction-before}\\
    =& \mathcal{P}(g)(x, f(x)^* k) + \mathcal{P}(f)(x, \Sigma_v  v^* \delta(v, g(x), k))\\
    =& \mathcal{P}(g)(x, f(x)^* k) + \mathcal{P}(f)(x, g(x)^* k)\label{eq:contraction-after},
\end{align}
which is the product rule for pullbacks (note the conjugates). From \cref{eq:contraction-before}
to \cref{eq:contraction-after}, the contraction met the delta function and
resulted in a substitution of $v$ into $g(x)$, and no polymorphic
object is present in the final result.

\subsection{Complex Numbers}%
\label{sub:Complex Numbers}

The primitive operations we listed in \cref{tab:primitive-lambdas} contain
mysterious complex conjugates. We now explain our treatment of complex numbers,
which leads to the complex conjugates in primitive pullbacks. The main
difficulty in dealing with complex numbers is that the standard complex analysis
does not prescribe a useful gradient for
optimization. For example, minimizing $z \mapsto |z|^2$ is evidently equivalent
to minimizing $(a, b) \mapsto a^2 + b^2$, but a Cauchy-Riemann argument shows
that $|z|^2$ is nowhere analytic, so the pullback makes no sense. For a real and scalar
valued function, this problem is partly resolved through the Wirtinger
derivative~\cite{gunning1965analytic}
\begin{equation}
    \partial f(z) / \partial z = \partial f(z) / \partial a + i \partial f(z) / \partial b, \quad z = a + i b, \label{eq:wirtinger}
\end{equation}
which can be used for, e.g., gradient descent. 

This formalism is insufficient for symbolic automation because it does not
handle the case where $f(z)$ is complex and requires splitting $z$ into
its real and imaginary parts. Differentiating a complex function $f(z)$ may seem unnecessary
when the objective function to optimize is always a real scalar. However, we
differentiate $f(z)$ by differentiating its constituents, which are complex-valued functions. Moreover, representing a complex gradient in terms of the real
and imaginary parts of $z$ is not acceptable for symbolic purposes, and it is
preferable to avoid splitting a complex variable to begin with (rather than trying
to reassemble them from the real and imaginary parts in the end).

These problems can be resolved by extending the definition of pullback to
complex numbers. We start by proposing the operators $\mathcal{V}$ and
$\mathcal{W}$
\begin{align}
    \mathcal{V}(z) &\triangleq
    \begin{bmatrix} \mathrm{Re}(z_1) &  \mathrm{Im}(z_1) & \ldots & \mathrm{Re}(z_n) & \mathrm{Im}(z_n) \end{bmatrix}^T,\\
    \mathcal{W}(f) &\triangleq v \mapsto \mathcal{V} (f(\mathcal{V}^{-1}(v))).
\end{align}
$\mathcal{V}$ and $\mathcal{V}^{-1}$ establish an isomorphism between
$\mathbb{C}^N$ and $\mathbb{R}^{2N}$ so that we can convert a complex problem
to a real one that is equivalent. Analogously, $\mathcal{W}$ converts between
$\mathbb{C}^N \to \mathbb{C}^M$ and $\mathbb{R}^{2N} \to \mathbb{R}^{2M}$. 
One can check that the following identities hold
\begin{align}
    \forall f \in \mathbb{C}^N \to \mathbb{C}^M,& \quad \mathcal{V}(f(z)) = (\mathcal{W}(f)) (\mathcal{V}(z)),\label{eq:vec-shove}\\
    \forall f \in \mathbb{C}^N \to \mathbb{R},& \quad f(z) = \mathcal{V}(1)^T \cdot (\mathcal{W}(f)) (\mathcal{V}(z)).
\end{align}

To minimize a scalar-valued function $f(z)$ over $z$, we can equivalently
minimize the real function $u \mapsto \mathcal{V}(1)^T \cdot (\mathcal{W}(f))(u)$
and convert $u$ to the corresponding complex number with $z =
\mathcal{V}^{-1}(u)$. The gradient of the real function can be written as $(\mathcal{J}(\mathcal{W}(f)))(u)^T \cdot
\mathcal{V}(1)$. Transforming this vector back into the complex space gives the
complex gradient $\mathcal{V}^{-1} \left((\mathcal{J} (\mathcal{W}(f)))(u)^T\cdot
\mathcal{V}(1)\right)$. Therefore, we write the Wirtinger gradient as
\begin{equation}
    \nabla f(z) \triangleq \mathcal{V}^{-1} (\mathcal{J} (\mathcal{W}(f))(\mathcal{V} (z))^T \cdot (\mathcal{V} (1))).
\end{equation}

To be able to find the gradient through the pullback as $\nabla f(z) =
\mathcal{P}(f)(z, 1)$, we suggest to define the complex pullbacks as
\begin{equation}
    \mathcal{P}(f) \triangleq (z, k) \mapsto \mathcal{V}^{-1} (\mathcal{J} (\mathcal{W}(f))(\mathcal{V}(z))^T \cdot \mathcal{V}(k)) \label{eq:complex-pullback}.
\end{equation}
Since the pullback remains a vector Jacobian product just like \cref{eq:real-pullback}, 
the $\mathbf{B}$ and $\mathbf{C}$ rules are not affected by the
change.
Therefore, the only modification to the theory is to 
derive the pullbacks of the univariate primitives listed in \cref{tab:primitive-lambdas}
using \cref{eq:complex-pullback} instead of \cref{eq:real-pullback}. As an
example, writing $z = x + iy$ and $k = a + ib$, the pullback of the complex
conjugate can be derived as
\begin{align}
    \mathcal{W}(z \mapsto z^*) &= (x, y) \mapsto (x, -y),\\
    \mathcal{P}(z \mapsto z^*) &= (z, k) \mapsto \mathcal{V}^{-1} \left(
    \begin{bmatrix}
        1 & 0 \\ 
        0 & -1
    \end{bmatrix}
    \cdot \begin{bmatrix}
        a\\
        b
    \end{bmatrix}\right) = (z, k) \mapsto k^*.
\end{align}

\subsection{Equivalence Graph}%

The two combinator rules introduced contractions and delta functions that have
to be combined into substitutions. This can be implemented in many ways, but we choose
to adapt a technique called the equivalence graph
(e-graph)~\cite{10.5555/909447}, which has found applications in compiler
optimization~\cite{10.1145/1594834.1480915, 2021-egg} and symbolic
math~\cite{Cheli2021}. We will use it primarily to combine contractions with
delta functions and perform a number of simplifications including combining
terms that are identical due to tensor symmetries. The details of the e-graph is
outside the scope of differentiation and will be covered in a separate manuscript.

\section{Results}%
\label{sec:Results}

We now show four examples of increasing difficulty that can be easily handled by
\texttt{CombDiff}. The simplest problem can only be symbolically automated with
dedicated rules for the matrix calculus, whereas the harder problems are
challenging even for experts to differentiate by hand. In our examples, the
domains of the functions and the ranges of the sums do not affect the
differentiation but complicate the presentation---so we will assume them to be
infinite. 

\subsection{Quadratic Functions}%
\label{sub:quadratic functions}

We introduce the syntax of \texttt{CombDiff} by differentiating the quadratic
form $x^{*}A x$ with $A \in \mathbb{C}^{N \times N}$ being Hermitian. Many symbolic systems can
handle this, but typically through a dedicated implementation of
the ``matrix calculus.'' Our system produces the same result without rules for matrices
or vectors except for post-differentiation optimization.
\begin{lstlisting}[caption=Quadratic functions,label=lst:linear-reg]
f, _ = @pct (A::Her) -> pullback((x::CV) -> 
    sum((i, j), x(i)'*A(i,j)*x(j)))

df = redux(vdiff(f), settings=symmetry_settings) |> blaserize
\end{lstlisting}
The syntax of \texttt{CombDiff} is identical to that of Julia since it is implemented as
a Julia DSL through its macro system, which gives a different semantic to same
syntax.  To obtain the derivative of $x \mapsto x^{*} A x$, one write it in
terms of its mathematical definition, as shown in \cref{lst:linear-reg}. The result
of the definition is a function \texttt{f} with a context, the latter of which
is discarded for this example. The \texttt{vdiff} function performs the
differentiation on \texttt{f} and produces the correct but unsimplified gradient, which the
\texttt{redux} function takes as input and simplifies. The
\texttt{blaserize} function then rewrites the result in terms of linear algebra
operations to improve readability and computational efficiency by leveraging BLAS. The final result is shown in
\cref{eq:output-quadratic}, whose latex code is generated by the system. 
\begin{align}\label{eq:output-quadratic}
   A &\mapsto x \mapsto 2.0 \cdot A\cdot x.
\end{align}
The factor of two is a consequence of the symmetry of $A$. The mechanism for encoding and
exploiting the symmetry will be illustrated in \cref{sub:Hartree Fock}.

This example, although simple, demonstrates many of the benefits of our
approach that persist into the complex examples: there is no need to 
complicate the implementation with partial evaluation, matrix calculus, or
memory mutations; the $2 A x$ is in a form that can be directly fed to BLAS,
but it would not be necessary  here because the evaluation is essentially free
as an intermediate of evaluating $x^* A x$ thanks to  the symmetry.

\subsection{Conjugate Gradient}%

The conjugate gradient (CG) algorithm~\cite{MethodsOfConjHesten1952,
NumericalLineaTrefet2022} is a widely used algorithm.  We give a conceptually simple
derivation of CG  by minimizing $R = x \mapsto \frac{1}{2} x^T A x - b^T x$,
whose stationary condition yields $Ax = b$. We consider a general optimization
step parametrized by the step sizes $x + \alpha(r + \beta p)$, where $x \in
\mathbb{R}^N$ is the current iterate, $p\in \mathbb{R}^N$ is the previous step
direction and $r \in \mathbb{R}^N$ is the current gradient. The residual as a
function of the parameters after taking the gradient step is $(\alpha, \beta)
\mapsto R(x + \alpha(r + \beta p))$, which we minimize with respect to $\alpha$ and
$\beta$.

In \cref{lst:cg}, we first differentiate the residual without substituting
the objective function $R$ to obtain an abstract theory that is generally
applicable. Then we show that replacing $R$ with the quadratic form gives
the CG coefficients.

\begin{lstlisting}[caption=Conjugate gradient,label=lst:cg]
f, _ = @pct (A::Sym, r::RV, p::RV, b::RV, x::RV) -> begin
    R = (x::RV) -> 
        sum((i, j), 0.5 * x(i) * A(i, j) * x(j)) - sum(i, x(i) * b(i))
    pullback((alpha::R, beta::R) -> 
        R((i::N) -> x(i) + alpha * (r(i) + beta * p(i))))
end;
blaserize(f)

df = vdiff(f) |> blaserize

df = redux(vdiff(eval_all(f)); settings=symmetry_settings) |> blaserize
\end{lstlisting}
The result of line 9 of \cref{lst:cg} is shown in \cref{eq:output-cg-1}, which
gives a vector of two components. This shows that  we can differentiate through
unknown functions as a consequence of avoiding the partial
evaluation.
\begin{equation}\label{eq:output-cg-1}
\begin{aligned}
&(A, r, p, b, x) \mapsto \mathrm{let}\\& \quad R = x \mapsto (-1.0\cdot x^{T}\cdot b+0.5\cdot x^{T}\cdot A\cdot x)\\&\quad (\alpha, \beta) \mapsto (\nabla (R)((\alpha \cdot (\beta \cdot p+r)+x))^{T}\cdot (\beta \cdot p+r), \\
&\quad\quad\quad\quad\quad\;\; \nabla (R)((\alpha \cdot (\beta \cdot p+r)+x))^{T}\cdot p\cdot \alpha)\\& \mathrm{end}
\end{aligned}
\end{equation}
If we write $p_k = r + \beta p$,  $\nabla R(\alpha p_k + x)^T \cdot p_k = 0$ has
the interpretation that the gradient at the next iterate should be orthogonal to
the current step direction. Combined with $\nabla R(\alpha p_k + x)^T \cdot p
\cdot \alpha = 0$, we have a nonlinear system of two equations for $\alpha$ and
$\beta$, the coefficients and thus the solutions of which depends on $R$.

Once we substitute the quadratic form for $R$ in line 11 of \cref{lst:cg},
the gradient reduces to 
\begin{equation}\label{eq:output-cg-2}
\begin{aligned}
&(A, r, p, b, x) \mapsto (\alpha, \beta) \mapsto \\
&\quad(((\beta \cdot p+r)^{T}\cdot A\cdot (\alpha \cdot (\beta \cdot p+r)+x)-1.0\cdot (\beta \cdot p+r)^{T}\cdot b), \\
&\quad \alpha\cdot (p^{T}\cdot A\cdot (\alpha \cdot (\beta \cdot p+r)+x)-1.0\cdot b^{T}\cdot p)).
\end{aligned}
\end{equation}
 Using the fact that the gradient $r = Ax - b$ is orthogonal to the previous step
direction $p$, the two nonlinear equations can be solved by hand to get
\begin{align}
    \alpha &= \frac{ p_k^T \cdot (b - Ax)}{p_k^T A p_k} = 
    -\frac{(r + \beta p)^T r}{p_k^T A p_k}
    = -\frac{r^T \cdot r}{p_k^T A p_k},\\
    \beta  &= \frac{(b^T - x^T \cdot A) \cdot p - \alpha r^T \cdot A \cdot p}{\alpha p^T A p} = \frac{r^T A p}{p^T A p},
\end{align}
which can be recognized as the parameters that produce the conjugate gradient
method. 

\subsection{Hartree Fock}%
\label{sub:Hartree Fock}

We now return to the HF problem motivating this work, which is a
contraction between a fourth order complex tensor $J \in \mathbb{C}^{N\times N\times N \times N}$ and a complex matrix $C \in \mathbb{C}^{N \times N_e}$
\begin{align}\label{eq:hartree-fock}
J \mapsto C \mapsto \Sigma_{i, j, p, q, r, s}C_{p, i}^{*} C_{r, j}^{*} J_{p, q, r, s} C_{s, j} C_{q, i}.
\end{align}
Similar to
how the gradient of $x^{*} A x$ can be simplified when $A$ is Hermitian,
the gradient of the HF energy can be simplified based on the symmetries of the
electron repulsion integral tensor $J$, which are encoded through the type
system. For example, the first symmetry in line 4 of \cref{lst:hartree-fock}
specifies that $J_{pqrs} = J_{qpsr}^*$, and the second symmetry specifies that
$J_{pqrs} = J_{rspq}$.
\begin{lstlisting}[caption=Hartree Fock,label=lst:hartree-fock]
_, ctx = @pct begin    
    @space T begin
        type = (N, N, N, N) -> C
        symmetries = (((2, 1, 4, 3), :conj), ((3, 4, 1, 2), :id))
    end
end;
f, _ = @pct _ ctx (J::T) -> pullback((C::CM) -> sum((i, j, p, q, r, s),
    C(p, i)' * C(q, i) * C(r, j)' * C(s, j) * J(p, q, r, s))); f
    
df = redux(vdiff(f); settings=symmetry_settings)

df |> get_body |> decompose |> pp |> simplify |> first
\end{lstlisting}
The gradient can be obtained as usual in line 10 of \cref{lst:hartree-fock}
and the result is 
\begin{equation}\label{eq:hf-gradient}
J \mapsto C \mapsto \left(\mathtt{d}, \mathtt{d}_{1}\right) \mapsto \left(\Sigma_{j, p, r, s}J_{r, s, p, \mathtt{d}}^{*} C_{s, j}^{*} C_{p, \mathtt{d}_{1}} C_{r, j}\right) 4.0.
\end{equation}
If one writes  $\hat{J}(\mathtt{d}, \mathtt{p}) = \Sigma_{r,s}
J^*_{r,s,\mathtt{p},\mathtt{d}} \Sigma_{j} C_{s,j}^* C_{r,j}$, then
\cref{eq:hf-gradient} can be written as a matrix multiplication $\hat{J} \cdot
C$, and $\hat{J}$ can be  recognized as the Coulomb contribution to the Fock
matrix~\cite{ComputationalPThijss2012,MathematicalInLinL2019}.  In terms of the
performance, this result is in the form that the tensor contraction engines
expect; the factor of $4$ represents a $4\times$ saving due to the symmetries,
but the evaluation is again free as an intermediate of evaluating the objective.
Getting the gradient for free currently requires manual effort, but this
is made possible by the readability of the result.

As an illustration of the differentiation process, we find the Hessian vector
product using lower level operations of the system. In line 12 of
\cref{lst:hartree-fock}, the gradient first goes through the \texttt{decompose} function,
which breaks down the map in \cref{eq:hf-gradient} in to the $\mathbf{B}$
and $\mathbf{C}$ combinators. \texttt{pp} then applies the $\mathbf{B}$ and
$\mathbf{C}$ rules to obtain the pullback, which encodes the Hessian vector
product. The result is then simplified into 
\begin{equation}
\begin{aligned}
\left(C, \mathtt{k}\right) \mapsto \left(\mathtt{a}_{2}, \mathtt{a}_{3}\right) \mapsto \left(\Sigma_{\mathtt{d}, \mathtt{d}_{1}, p, s}C_{p, \mathtt{d}_{1}}^{*} J_{\mathtt{a}_{2}, s, p, \mathtt{d}} \mathtt{k}_{\mathtt{d}, \mathtt{d}_{1}} C_{s, \mathtt{a}_{3}}+\right.\label{eq:hf-hvp}\\
\Sigma_{\mathtt{d}, \mathtt{d}_{1}, p, r}J_{r, \mathtt{a}_{2}, p, \mathtt{d}}^{*} \mathtt{k}_{\mathtt{d}, \mathtt{d}_{1}}^{*} C_{p, \mathtt{d}_{1}} C_{r, \mathtt{a}_{3}}+ \\
\left. \Sigma_{\mathtt{d}, j, r, s}C_{r, j}^{*} J_{r, s, \mathtt{a}_{2}, \mathtt{d}} \mathtt{k}_{\mathtt{d}, \mathtt{a}_{3}} C_{s, j}\right) 4.0. 
\end{aligned}
\end{equation}

\subsection{Localized Wannier Functions}%

Lastly, we examine the theory of localized Wannier
functions~\cite{MaximallyLocalMarzar1997,marzari2012maximally}. The gradient used in the standard
codebase \texttt{Wannier90} is technically the projected gradient~\cite{MaximallyLocalMarzar1997}, which is partly the reason why \texttt{Wannier90} 
requires a dedicated optimizer. A derivation that produces the usual gradient can
be found in~\cite{PhysRevB.72.125119}. 

The optimization objective for this problem is a complex tensor contraction similar
to the HF energy
\begin{align}
    U &\mapsto \Sigma_{n, b} w(b) |\hat{\rho}(n, b)|^2, \\ \hat{\rho}(n, b) &=
    \Sigma_{k,p,q} U(p, n, k)^{*} S(p, q, k, k+b) U(q, n, k+b),
\end{align}
where $S \in \mathbb{C}^{N \times N \times N_k \times N_k}$ is known as the
overlap matrix and $U \in \mathbb{C}^{N \times N \times N_k}$ is known as the gauge.
Similar to the HF case, $S$ has a symmetry $S(p,q,k,k+b) = S(q,p,k+b,k)^*$,
which has to be leveraged to simplify the result. The difference from the HF
problem is that the index $k$ is on a periodic domain; one has to consider
arithmetic in the indices, and $b$ is on a symmetric domain.

For our system to fully simplify the gradient, all of the above assumptions have
to be encoded in the types, and these assumptions themselves are subtle enough
that one needs an expert to identify them. One can proceed without these
assumptions and get a suboptimal but nevertheless correct gradient, but we will
encode these assumptions in \cref{lst:loc-wannier} to show that this is
possible in principle.
\begin{lstlisting}[caption=Localized Wannier functions,label=lst:loc-wannier]
_, ctx = @pct begin
    @domain BZ begin
        base = I
        periodic = true
    end

    @domain X begin
        symmetric = true
        contractable = false
    end

    @space Mmn begin
        type = (N, N, BZ, BZ) -> C
        symmetries = (((2, 1, 4, 3), :conj),)
    end

    @space SV begin
        type = (I,) -> C
        symmetries = (((1,), :ineg),)
    end

    @space Gauge begin
        type = (N, N, BZ) -> C
    end
end;
f, _ = @pct _ ctx (S::Mmn, w::SV) -> pullback((U::Gauge) -> 
    begin
        rho = (n::N, b::X) -> sum((k::BZ, p, q), U(p, n, k)' * 
            S(p, q, k, k + b) * U(q, n, k + b))
        sum((n, b::X), w(b) * rho(n, b)' * rho(n, b))
    end); f
    
df = redux(vdiff(eval_all(f)); settings=symmetry_settings)
\end{lstlisting}
The result of the code is 
\begin{equation}
    \begin{aligned}
\left(S, \right.&\left. w\right) \mapsto U \mapsto \left(\mathtt{d}, \mathtt{d}_{1}, \mathtt{d}_{2}\right) \mapsto 4.0\cdot \left(\Sigma_{b,p,k,\mathtt{p},q}U\left(\mathtt{p}, \mathtt{d}_{1}, \left(b+k\right)\right)^{*}\cdot w\left(b\right)\cdot\right. \\
&\left. U\left(q, \mathtt{d}_{1}, k\right)\cdot U\left(p, \mathtt{d}_{1}, \left(\mathtt{d}_{2}+b\right)\right)\cdot S\left(\mathtt{d}, p, \mathtt{d}_{2}, \left(\mathtt{d}_{2}+b\right)\right)\cdot S\left(\mathtt{p}, q, \left(b+k\right), k\right)\right).
    \end{aligned}
\end{equation}
This result requires one further step of simplification to be recognized as
similar to the gradient derived by hand in~\cite{AccuratePolariStenge2006}
\begin{equation}
\begin{aligned}
\left(S, \right.&\left. w\right) \mapsto U \mapsto \left(\mathtt{d}, \mathtt{d}_{1}, k\right) \mapsto 4.0
\cdot \left(\Sigma_{b,p,q} w\left(b\right)\cdot \hat{\rho}(\mathtt{d}_1, b)^*\right. \\
&\left. U\left(p, \mathtt{d}_{1}, \left(k+b\right)\right)\cdot S\left(\mathtt{d}, p, k, \left(k+b\right)\right)\right).
\end{aligned}
\end{equation}
Once again, the gradient is free as a byproduct of evaluating the objective function,
although  it is more subtle to see. However, due to the arithmetics on the
indices,  neither the objective nor the gradient can be fed directly into a
tensor contraction engine.

\section{Conclusion}%
\label{sec:Conclusion}

We have developed a novel framework for automating variational differentiation
based on the combinatory logic. We proposed a pair of differentiation rules for
the $\mathbf{B}$ and $\mathbf{C}$ combinators, which spans the space of
computable functions. Our differentiation framework is then completed with a
generalization to complex numbers through a redefinition of the pullbacks. Our
approach enables VD and backpropagation while staying free from any numerical
evaluations. Consequently, we are able to  keep the gradients human readable,
minimize code complexity, and avoid performance losses inherent to conflating
differentiating with performance optimizations.


The main theoretical limitation of our method is the performance of the e-graph search when
tensor symmetries are involved. Although the hardest example in the paper (MLWF)
requires less than ten seconds on a M1 Pro Processor, which is acceptable for
interactive use, the time complexity can grow rapidly with larger problems. This
issue can be mitigated by guiding the search on the graph with
the compressibility of the ASTs, which also has the benefit of improving
the numerical performance.





\section*{Acknowledgements}
K.L. and A.D. were supported by the SciAI Center, and funded by the Office of Naval Research (ONR), under Grant Number N00014-23-1-2729.
We like to thank Dr. Guido Falk von Rudorff for the discussion on differentiating the stationary points.

\bibliographystyle{siamplain}
\bibliography{lib}

\end{document}